\documentclass[preprint,12pt]{elsarticle}

\usepackage{amssymb}
\usepackage{mathtools}
\usepackage{booktabs}
\usepackage{hyperref}

\usepackage{graphicx}
\usepackage[caption = false]{subfig}
\usepackage{color}
\journal{Journal of Computational Physics}

\begin{document}

\pagenumbering{gobble}

{
\setlength{\parindent}{0pt}
\setlength{\parskip}{0.5cm plus4mm minus3mm}

\textbf{Electronic coarse graining enhances the predictive power of molecular simulation allowing challenges in water physics to be addressed} by Flaviu S. Cipcigan, Vlad P. Sokhan, Jason Crain and Glenn J. Martyna

Preprint accepted in the Journal of Computational Physics. \\
\url{dx.doi.org/10.1016/j.jcp.2016.08.030}

}

\newpage
\pagenumbering{arabic}

\begin{frontmatter}



\title{Electronic coarse graining enhances the predictive power of 
       molecular simulation allowing challenges in water physics to be addressed}

\author[edi,npl]{Flaviu S. Cipcigan}
\author[npl]{Vlad P. Sokhan}
\author[edi,npl]{Jason Crain}
\author[ibm]{Glenn J. Martyna}

\address[edi]{School of Physics and Astronomy, University of Edinburgh, Peter Guthrie Tait Road, Edinburgh EH9 3FD, United Kingdom}
\address[npl]{National Physical Laboratory, Hampton Road, Teddington, Middlesex TW11 0LW, United Kingdom}
\address[ibm]{IBM T. J. Watson Research Center, Yorktown Heights, New York 10598, USA}

\begin{abstract}
One key factor that limits
the predictive power of molecular dynamics simulations is the accuracy and transferability
of the input force field. Force fields are challenged by heterogeneous
environments, where electronic responses give rise to biologically important forces 
such as many-body polarisation and dispersion.
The importance of polarisation was recognised early-on
and described by Cochran in 1959 \cite{Cochran1959, Sangster1976}.
However, dispersion forces
are still treated at the two-body level and in the dipole limit, although the 
importance of three-body terms in the condensed phase was demonstrated 
by Barker in the 1980s \cite{Barker1986a, Barker1986b}.
A way of treating both polarisation and dispersion
on an equal basis is to coarse grain the electrons a molecular moiety to a single 
quantum harmonic oscillator, as suggested as early as the 1960s 
by Hirschfelder, Curtiss and Bird \cite{HCB}.
This treatment, when solved in the strong coupling limit, gives all orders of 
long-range forces. In the last decade, the tools necessary to 
exploit this strong coupling limit 
have been developed, culminating in a transferable model of water 
with excellent predictive power across the  
phase diagram. This transferability arises since the environment 
identifies the form of long range interactions, rather than the 
expressions selected by the modeller. 
Here, we discuss the role of electronic coarse-graining in 
predictive multiscale materials modelling and describe the first implementation 
of the method in a general purpose molecular dynamics software, QDO\_MD.
\end{abstract}


\end{frontmatter}



\section{Introduction}

\begin{figure}[t]
  \includegraphics[width=\textwidth]{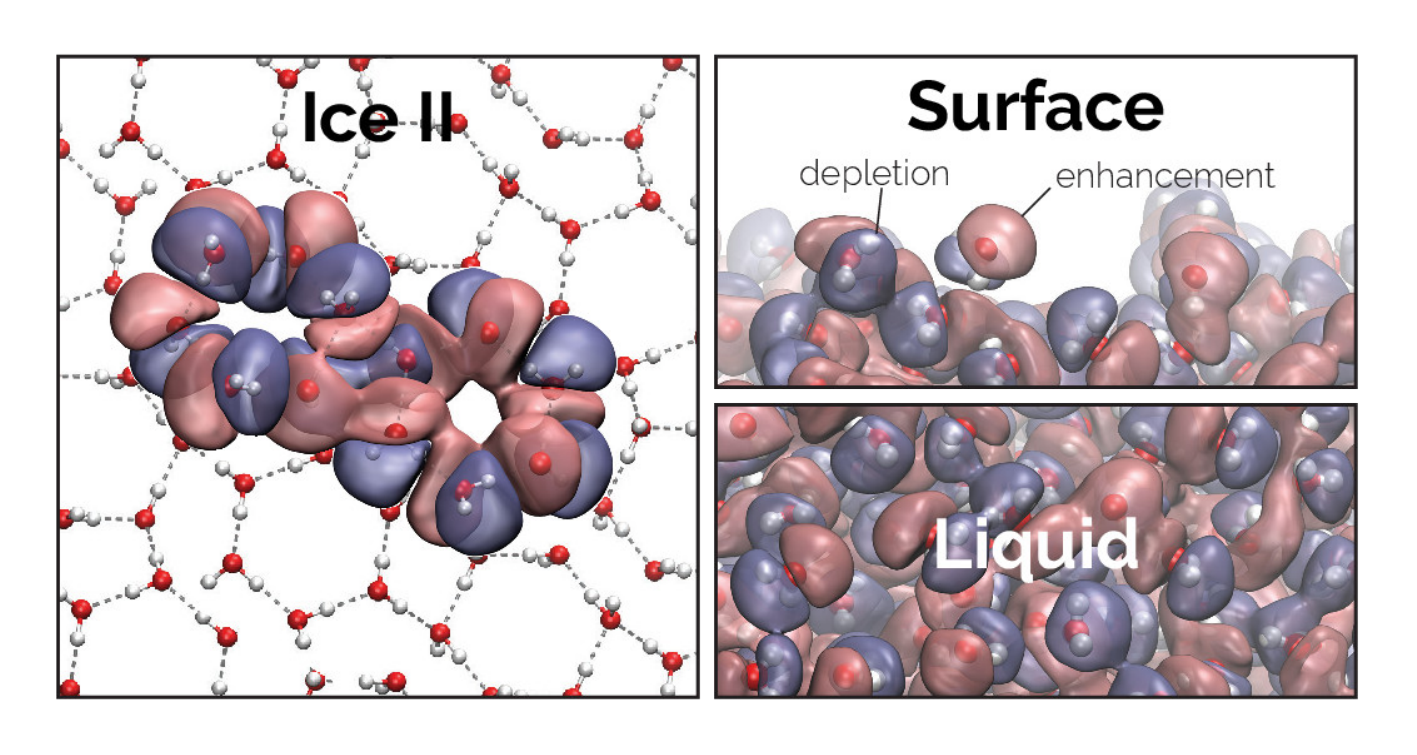}
  \caption{Examples of environments where the properties of QDO-water
           have been studied and shown to match experiment: ice II, ambient
           temperature liquid and the surface of the liquid.
           The images illustrate the electronic responses 
           of individual molecules, with red and blue isosurfaces 
           corresponding to regions of enhancement and depletion of 
           electronic density, respectively.} 
  \label{fig:environments}
\end{figure}

Molecular modelling has become an integral part of research in all areas of condensed 
matter, forming a new part of the scientific method that elucidates 
principles and accelerates the discovery of 
new materials \cite{Schames2004, Durrant2011}. This computational revolution
has been driven by the exponential scaling
of computer hardware \cite{Frank2001, Dennard1999} along with the implementation of scalable
software on these platforms \cite{Phillips2005, Tuckerman2000, Bohm2008, Vadali2004}.
An important part of the equation remains to improve the accuracy and transferability
of the employed force fields, which ultimately define the accuracy of results and
lay the basis for predictive modelling, given sufficient sampling of phase space.

Standard force fields, designed to be numerically efficient, are 
created by fitting a subset of thermodynamic and
structural properties to an analytic expression for the potential energy, 
typically consisting of sums of
pairwise terms \cite{Rapaport2004}. In many systems, this approach has proved successful
even if the accuracy of such force fields is as best around 1 kcal/mol.
For example,
Schames et al. \cite{Schames2004} used molecular dynamics simulations to discover
 a hidden binding site
in HIV integrase, a molecule that enables the integration of HIV's 
genetic material into the host cell DNA. This discovery led to experimental confirmation
\cite{Hazuda2004} and subsequent development 
of raltegravir, a successful medicine that halts the progression of HIV into AIDS.
Another success was the simultaneous discovery by simulation \cite{Offman2010}
and experiment \cite{Wei2010}
of the molecular basis of mutations in a gene responsible for Gaucher's disease,
a genetic disease causing the build-up of fatty substances in the organism.

Nonetheless, strategies based on fitting to a fixed functional are not guaranteed to be 
transferable to heterogeneous environments far
from the region of parametrisation. One such environment is, for example, 
a bacterial membrane penetrated by a peptide. 
The shape of the peptide inside the membrane depends strongly 
on the force fields used \cite{Wang2014}, with only experiments
currently being able to distinguish between possibilities.

Here we present our recent development of a force field for materials modelling 
\cite{Sokhan2015a, Sokhan2015b, Cipcigan2015, Jones2013b, Jones2013a, Whitfield2006, Whitfield2007}
that is transferable, physically motivated, and implemented in
the general purpose molecular dynamics software QDO\_MD. The technique treats
many-body polarisation 
and many-body dispersion on the same footing by representing 
electronic distributions of individual atoms and 
molecular moieties using a single coarse grained particle. 
This particle, known as a Quantum Drude Oscillator \cite{Whitfield2006,Whitfield2007} (QDO),
consists of a negative charge bound to
a positive centre by a quantum harmonic oscillator. 
QDOs generate many-body polarization, 
many-body dispersion interactions and cross terms beyond the dipole limit \cite{Jones2013c}
and represent a non-perturbative approach that generates all long-range forces
within Gaussian statistics when sampled in strong coupling.

The QDO approach has been employed to construct a general purpose water
model (QDO-water) that is transferable without fitting to any condensed
phase data (neglecting bond breaking and making).
QDO-water consists of a rigid molecular geometry (made out of
fixed charges), short-range empirical repulsion and a QDO. The model is
parametrised using the static multiple moments of an isolated molecule,
its polarisability, the dipole-dipole dispersion coefficient ($C_6$)
and the energy landscape of the dimer. 
Its properties have been studied
in a diverse set of environments (depicted in Fig.~\ref{fig:environments}), 
ranging from high-pressure ice (ice II),
liquid water, steam, the liquid-vapour interface, and supercritical water. 
The model predicts density, surface tension,
enthalpy of vaporisation and dielectric constant across these wide range of 
environments, demonstrating a formerly unrealized level of transferability.

\section{Predictive materials modelling using electronically coarse grained methods}

\begin{figure}[t]
  \includegraphics[width=\textwidth]{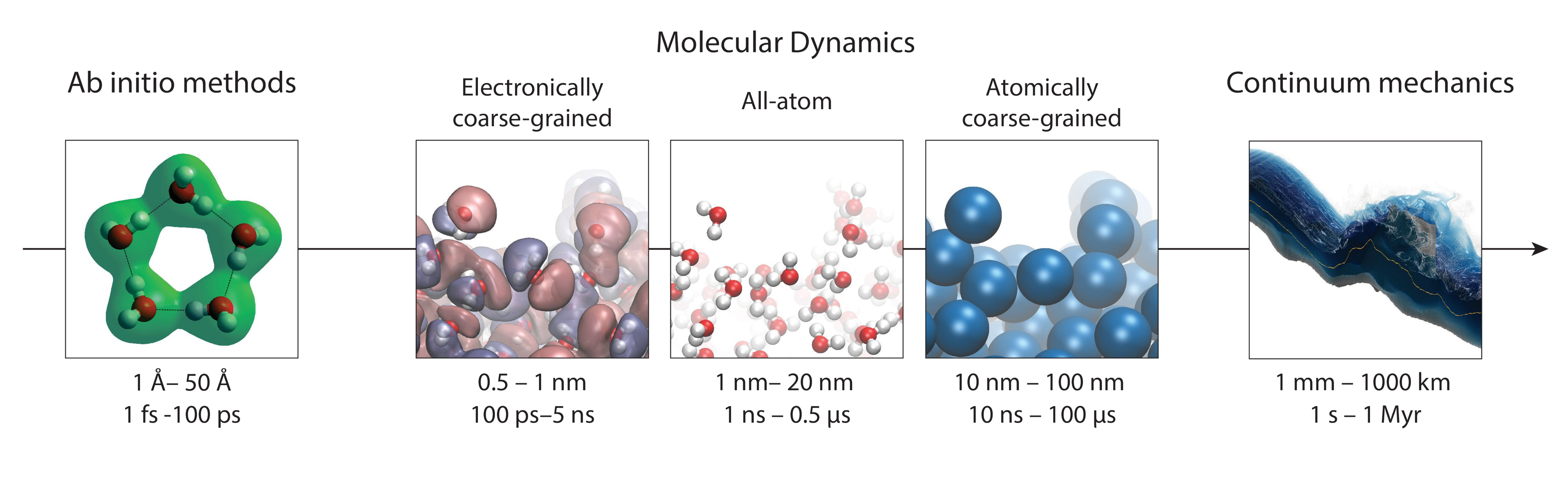}
  \caption{Various simulation techniques arranged according to the timescales and lengthscales
            which they sample. The left-most image, originally by Chaplin
           \cite{Chaplin}, represents electron densities of the water pentamer obtained via
           ab initio calculations. The next is the electronically coarse grained 
           representation of the air-water surface, 
           where red and blue isosurfaces depict regions of enhancement
           and depletion of electronic density. This is followed by a classical 
           molecular dynamics simulation and 
           an atomically coarse grained study of the same interface.
           The right-most image represents a particle-based realtime
           simulation of water flow over complex landscape using a method developed by Chentanez
           and M{\"u}ller \cite{Chentanez2011}.
           The left and right-most images are are used with permission from their respective authors
           while the rest are original work by the authors of this paper.} 
  \label{fig:multiscale}
\end{figure}

Fig.~\ref{fig:multiscale} shows where the QDO method stands on the spectrum
of modelling techniques, occupying a previously vacant place 
between ab initio methods and all-atom molecular dynamics.
QDOs' guiding philosophy is ``instead of treating an exact system in an approximate
way, treat an approximate system in an exact way''.

In a physical sense, the QDO model is solvable in strong coupling, allowing for 
an approach that contains all long-range, many-body interactions to all orders. 
The lack of truncation means that the modeller does not pick the symmetry by choice 
of truncation but rather the environment picks the key terms. This type of simplified 
but rich model is at the heart of modern theories because its lack of bias allows 
the essential physics to emerge naturally.

The price paid for the richness of the QDO model at low computational cost (order N) 
is the approximation made in coarse graining: assuming localized electrons in an 
insulator respond according to Gaussian statistics and the neglect of exchange 
effects, which are assumed to be short range. This is consistent with the general 
spirit of coarse grained approaches where exactness is sacrificed while maintaining 
sufficient accuracy for the problem at hand.  Basically, QDOs are a coarse graining of 
a full high level electronic structure due to their lack of truncation, 
while standard force fields are a coarse graining of QDOs, with a truncation to dipole 
limit pair-wise dispersion and a neglect of polarization or its inclusion 
at the many-body dipole level. As demonstrated in the next section,
systems of single QDOs per atom or molecular moiety solved in strong coupling
are sufficient
to model molecules that are isotropic (such as noble gases) or close to isotropic (such as
simple hydrides like methane and water). To model anisotropic responses, multiple QDOs can
be pinned to individual points in the molecular frame \cite{Stone2007}.

Water is an ideal system in which to illustrate the importance of a complete treatment of
electronic responses. Water's
structure arises due to a competition between directional hydrogen bonds, which
favour an open ice-like local structure and van der Waals forces, which favour a close-packed
local structure. Many-body polarization and dispersion interaction are key to 
distinguishing the dominant motifs \emph{under the conditions of interest}.  

Hydrogen bonds are known to be cooperative, meaning that their interaction
strength changes depending on environment thus leading to 
a wide variety of motifs emerging.
A simple reporter of this cooperation is the molecular
dipole moment, which changes from a value of 1.85 D in the gas phase to an estimated value of 2.5--3.6 D
in liquid \cite{Jones2013a, Kemp2008}, where four hydrogen bonded motifs are dominant. This sensitivity
to local structure can be used as a fingerprint of various local motifs, as we have reported
in reference \cite{Sokhan2015b}. There, we showed that a cusp in the dipole moment as a function of
temperature coincides with experimental maxima in heat capacity,
separating the supercritical region in ``gas-like'' and ``liquid-like'' regions 
\cite{Simeoni2010} and revealing the
liquid-gas Widom line \cite{Widom1972} at a molecular level.
We also showed that such cooperation
makes  hydrogen bonds more asymmetric than previously thought \cite{Cipcigan2015}, 
with a molecule favouring the 
loss of an acceptor bond
(on the side of the oxygen) over that of a donor bond (on the side of the hydrogen)
in both the liquid phase and at the liquid--gas interface.

Dispersion interactions are also important in water. Including these responses is
essential to generate even the basic structure of water at room temperature. 
 An illustrative example of this balance is the overstructuring of 
room-temperature water by DFT \cite{Lin2012}, where including the electron correlations that lead
to accurate van der Waals interactions is challenging
(although, there have been promising results 
using a similar technique of embedded quantum harmonic oscillators at the dipole level
\cite{Tkatchenko2012}). 
In our water model the dispersion coefficients are tunable, while keeping the polarisation fixed,
by changing the parameters of the QDO. In Ref. \cite{Jones2013a}, we showed that by  changing
the strength of van der Waals interactions, the density of liquid water can be changed by 15\% while
still keeping a hydrogen bonded local structure. 

In addition, three-body effects account for as much as
25\% of the binding energy of water, according to the estimates of Keutsch et al. \cite{Keutsch2003}. 
Thus, monoatomic
two-body potentials for the water molecule can only reproduce water's condensed phase properties
if their parameters are allowed to vary with state point \cite{Johnson2007, Chaimovich2009}. 
Even a full-atom description of 
water cannot be transferable if
its electrostatics is fixed. Vega and Abascal \cite{Vega2005} showed that a nonpolarisable
model of water cannot simultaneously be fit to the melting temperature and temperature of
maximum density.

Therefore, in order to create a truly transferable energy surface for molecular simulations,
which is therefore predictive outside the regime of parametrisation, the 
electronic effects need to be taken into account since they give rise 
to non-trivial many-body polarization and dispersion. QDOs do
so while keeping efficiency close to that of all-atom molecular dynamics and scaling
as $N \log{N}$ with the number of atoms $N$. This strategy
can be used to create models that are parametrised without any condensed phase input, yet
generate the properties of condensed phases. 

\section{Electronic coarse graining using Quantum Drude Oscillators}
\subsection{Dispersion interactions between quantum harmonic oscillators}

In order to gain
intuition into how quantum harmonic oscillators replicate electronic responses,
consider a simple model \cite{Hirschfelder1964}: a negative
charge $-q$ of mass $m$, free to move in a dimension $x$. The charge is
localised by connecting it by a spring of frequency $\omega$ to a 
positive charge $+q$ fixed at the origin. 

Consider adding a positive, fixed test charge $Q$ at a distance $R >> x$ from the 
origin. The system is in equilibrium when the Coulomb force $E q$ on the 
negative charge equals the spring force $m \omega^2 x$. Balancing the two forces results in the
oscillator acquiring a dipole moment $\mu = q x = \frac{q^2 E}{m \omega^2}$ and thus having a 
dipole polarisability defined as:
\begin{equation}
\alpha_1 := \frac{\mu}{E} = \frac{q^2}{m \omega^2}.
\end{equation}

In order to account for dispersion effects, the system has to be treated quantum 
mechanically.  Consider two identical 
oscillators separated by a distance $R$ interacting via a term $c(R)$. The
leading order interaction is dipole-dipole, meaning that the term $c(R)$ is proportional to
$R^{-3}$. Thus, the Hamiltonian is:
\begin{equation}
\hat{H} = \frac{\hat{p}_1^2}{2 m} + \frac{\hat{p}_2^2}{2 m} + \frac{1}{2} m \omega^2 \left( x_1^2 + x_2^2 + c(R) \, x_1 x_2 \right)\,.
\end{equation}
Changing coordinates to $\hat{\rho}_\pm = \frac{1}{\sqrt{2}} \left(\hat{p}_1 \pm \hat{p}_2\right)$ and 
$\xi_\pm=\frac{1}{\sqrt{2}} \left( x_1 \pm x_2 \right)$ decouples the Hamiltonian into two independent
harmonic oscillators of frequency $\omega \sqrt{1 \pm c}$.
\begin{equation}
\hat{H} = \left( \frac{\hat{p}_+^2}{2 m} + \frac{1}{2} m \omega^2 \left(1 + c\right) \xi_+^2 \right)
        + \left( \frac{\hat{p}_-^2}{2 m} + \frac{1}{2} m \omega^2 \left(1 - c\right) \xi_-^2 \right)
\end{equation}

The ground state energy of the coupled system is hence given by, to leading order in $c$:
\begin{equation}
\begin{split}
E_0 &= \frac{1}{2} \hbar \omega \left(\sqrt{1 + c} + \sqrt{1 - c} \right) \\
    &\approx \hbar \omega \left( 1 - \frac{1}{8} c^2 + O(c^3)\right).
\end{split}
\end{equation}

\begin{figure}[htp]
\centering
  \includegraphics[width=1.0 \textwidth]{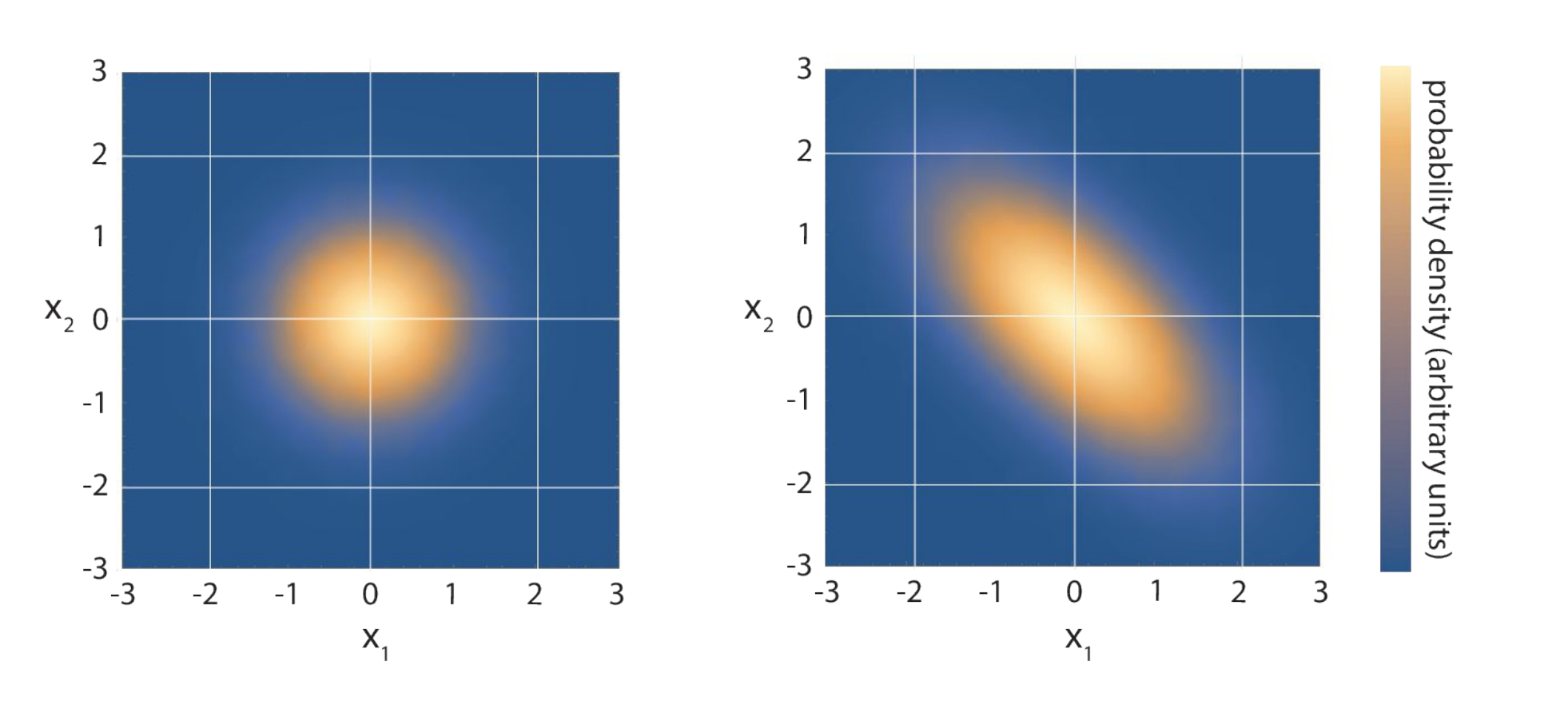}
\caption{The ground state probability density of two one dimensional quantum harmonic oscillators interacting 
         via a $c(R) x_1 x_2$ term. The left
         panel illustrates the non-interacting case, with $c(R) = 0$. The right panel 
         illustrates the interacting case, with $c(R) > 0 $. The spread of the
         wavefunction across the diagonal represents the electron correlation the gives rise
         to attractive van der Waals forces in real systems.}
\label{fig:1d_oscillator_ground}
\end{figure}

Fig.~\ref{fig:1d_oscillator_ground} shows the ground state probability distribution
$\psi^2(x_1, x_2)$ in the independent ($c=0$) and correlated ($c>0$) cases. When
$c$ increases, it becomes more probable for $x_1$ 
and $x_2$ to have the opposite sign and less probable for them to have the same sign. Since
the energy is proportional to $\int \, dx_1 \, dx_2 \, \psi^2(x_1, x_2) x_1 x_2 c(R)$, this correlation leads
to the attractive force between two neutral molecules proportional to $-c(R)^2 \sim -R^{-6}$,
just as the leading order term in the van der Waals potential.

This analysis shows that a one dimensional quantum harmonic oscillator can capture 
two basic effects of long range forces: polarisation and dispersion. In this case,
these interactions are limited to dipole-limit forces. To capture the full set of
interactions one has to consider the full three dimensional system.

%=======================================================
\subsection{Full quantum model}

A Quantum Drude Oscillator (QDO) \cite{Whitfield2006} is made out of a light, negative particle of
charge $q$ connected 
to a heavy positive \emph{centre} by a harmonic spring of frequency $\omega$. The bound state
of this system is a \emph{drudon}, with reduced mass $m$ and Hamiltonian
\begin{equation}
\hat{H} = \frac{\hat{\mathbf{p}}^2}{2 m} + \frac{1}{2} m \omega^2 \mathbf{x}^2.
\end{equation}

Perturbing this Hamiltonian via a point charge $Q$ at a distance $R$ gives
a first order correction to the energy of zero and a second order correction of \cite{Jones2013c}:
\begin{equation}
E^{(2)} = - \sum_{l=0}^{\infty} \frac{Q^2 \alpha_l}{2 R^{2 l + 2}},
\end{equation}
where the polarisabilities $\alpha_l$ are given by
\begin{equation}
\alpha_l = \left[ \frac{q^2}{m \omega^2} \right] 
\left[ \frac{(2 l - 1)!!}{l}\right] \left[ \frac{\hbar}{2 m \omega} \right]^{l-1}.
\end{equation}

Considering two QDOs separated by a distance $R$ gives a
first order correction to the energy of zero and a second order correction of \cite{Jones2013c, Fontana1961}:
\begin{equation}
E^{(2)} = - \sum_{n=3}^{\infty} C_{2n} R^{-2n}
\end{equation}
with
\begin{equation}
\begin{split}
C_6 &= \frac{3}{4} \alpha_1 \alpha_1 \hbar \omega, \\
C_8 &= 5 \alpha_1 \alpha_2 \hbar \omega, \\
\ldots
\end{split}
\end{equation}

Therefore, the full model has a 
complex and rich set of long range responses, which are replicated
with only three parameters. This means the responses are
correlated and thus one needs to check whether these correlations
are satisfied in real atoms and molecules.

\subsection{Invariant relationships between response coefficients}

As the responses of QDOs depend only on three parameters, these parameters
can be eliminated, resulting in the following
invariants that can be used to verify the accuracy of the QDO approximation \cite{Jones2013c}:
\begin{equation}
\begin{split}
\sqrt{\frac{20}{9}}\frac{\alpha_2}{\sqrt{\alpha_1 \alpha_3}} &= 1, \\
\sqrt{\frac{49}{40}}\frac{C_8}{\sqrt{C_6 C_{10}}} = 1, \\
\frac{C_6 \alpha_1}{4 C_9}=1.
\end{split}
\end{equation}

\begin{figure}[htp]
\centering
\includegraphics[width=\textwidth]{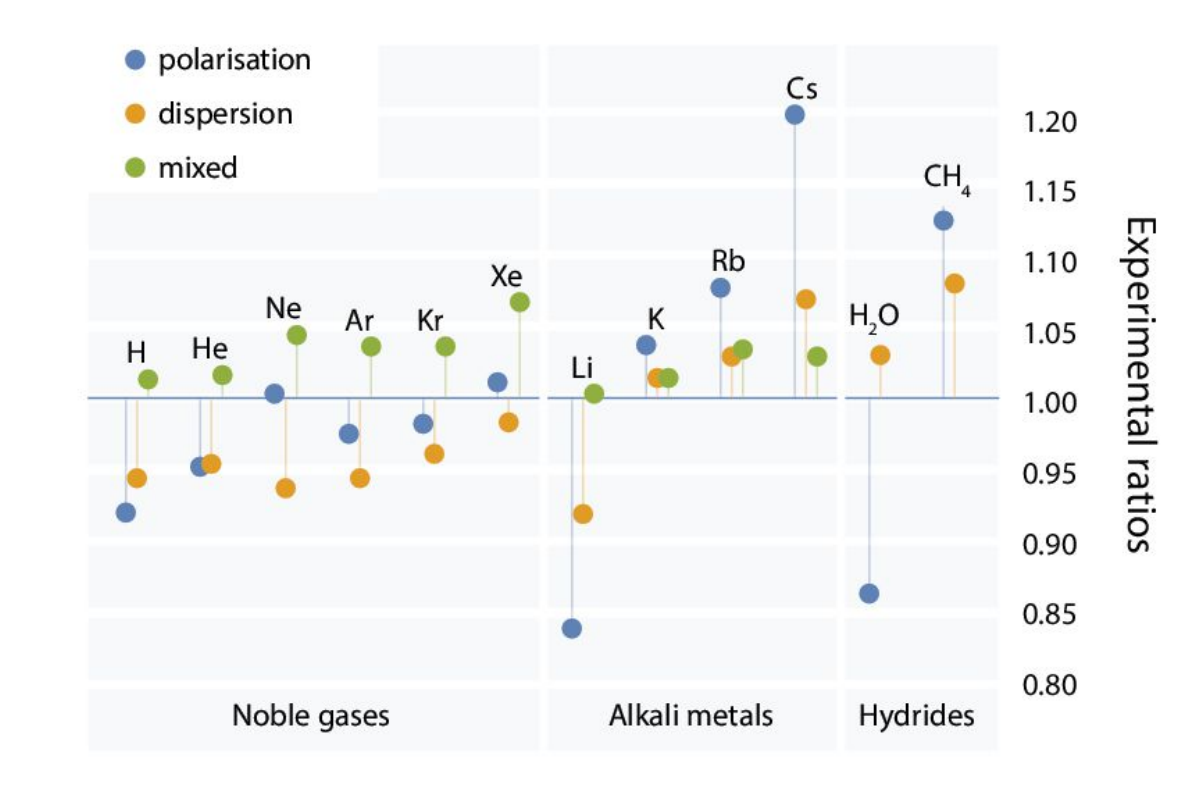}
\caption{Three types of invariant ratios between polarisation and dispersion coefficients,
         predicted by Quantum Drude Oscillators. Polarisation ratios involve only polarisation
         coefficients and analogously for dispersion ratios. Mixed ratios involve both polarisation
         and dispersion coefficients. Deviation from theory is shown for three types of atoms and molecules: noble 
         gases, alkali metals and small hydrides.}
\label{fig:invariants}
\end{figure}

Comparing these ratios with the responses of real molecules gives
the results shown
in Fig.~\ref{fig:invariants} (reproduced from Ref.~\cite{Jones2013c}).

The agreement is good for the noble gases, with ratios within 10\% of experiment,
as their shells are fully filled and
thus the distribution of electrons is nearly spherical. The agreement is also
within 10\% for most alkali metals, for which the last electron contributes mostly to polarisation.
Small hydrides such as water and methane  agree within 15\%. This agreement is due to the electronegativity
of atoms such as oxygen, which centres most of the electronic charge on them thus
resulting in charge distributions and responses that are close to spherical.

\subsection{Parametrisation}

Aside invariants, the responses of a QDO can be inverted to give 
its properties as a function of the responses. The relevant relations, 
which are used to parametrise a QDO are as follows:

\begin{gather}
\omega=\frac{1}{\hbar} \frac{4 C_6}{3 \alpha_1^2}, \\
m=\frac{\hbar}{\omega} \frac{3 \alpha_1}{4 \alpha_2} \quad \textrm{or} \quad m=\frac{\hbar}{\omega} \frac{5 C_6}{C_8}, \\
q=- \sqrt{m \omega^2 \alpha_1}.
\label{eq:fitting}
\end{gather}

%=======================================================
\section{Simulating Quantum Drude Oscillators via two temperature path integral molecular dynamics}

Path integral molecular dynamics, shortened to PIMD, is a method of sampling a quantum
mechanical partition function such as that of a QDO, using  classical methods 
\cite{Parrinello1984, Tuckerman1993}.
This section highlights its basic derivation, with the full, QDO-based method being described
in Refs.~\cite{Whitfield2007, Jones2013b}.

Consider a quantum
mechanical system with degrees of freedom $\vec{x}$ whose state is described by
a density matrix $\rho(\vec{x}, \vec{x}^\prime)$. The partition function of 
the system is \cite{Feynman1998}:
\begin{equation}
Z(\beta) = e^{-\beta \hat{H}} = \int \mathbf{d}\vec{x} \, \rho(\vec{x}, \vec{x}; \beta),
\end{equation}
where $\mathbf{d}\vec{x}$ represents a small volume element in the 
$\mathrm{dim}(\vec{x})$-dimensional space spanned by the system's degrees of freedom.

Since the density matrix is an exponential, it can be factorised into higher temperature
(smaller $\beta$)
slices to give the following path integral representation,
\begin{equation}
  Z(\beta) =  \prod_{i=1}^P \int \mathbf{d}\vec{x}_i \, \rho(\vec{x}_i, \vec{x}_{i+1}; \beta / P).
\end{equation}

The reason for this factorisation is the fact that increasing the temperature, thus
decreasing $\beta$ makes density matrices more classical and thus easier to approximate.
The approximation used comes from a
Trotter factorisation \cite{Whitfield2007, Trotter1959}, where the Hamiltonian is split 
into a reference $\hat{H}_0$,
with known density matrix  $\rho_0 = e^{-\tau \hat{H}_0}$
and a perturbation $V(\vec{x})$ as follows:
\begin{equation}
  \rho(\vec{x}, \vec{x}^\prime; \tau) \approx 
    e^{-\tau V(\vec{x}) / 2} 
    \rho_0(\vec{x}, \vec{x}^\prime; \tau) 
    e^{-\tau V(\vec{x}^\prime) / 2} + O(\tau^3).
\end{equation}

%$V(\vec{x})$ is the external potential, which typically includes point charges and other QDOs. 
% This gives the following reference density matrix:

In the case of QDOs,  $H_0$ is the Hamiltonian of the isolated system, with the following reference 
density matrix:
\begin{equation}
  \rho_0(\vec{x}_i, \vec{x}_{i+1}; \tau) = 
    \left[
      \frac{\alpha_P(\tau)}{\pi}  
    \right]
    \exp\left[
      \alpha_P(\tau) (\vec{x} - \vec{x}^\prime)^2 + \frac{\lambda_P(\tau)}{2}(\vec{x}^2 + (\vec{x}^\prime)^2)
    \right].
\end{equation}
The coefficients $\alpha_P(\tau)$ and $\lambda_P(\tau)$ are defined in \ref{app:coeff}.

Note that this form introduces strong nearest-neighbour coupling between the coordinates $\vec{x}_i$.
To remove this coupling, the density matrix is diagonalised to the independent coordinates $\vec{u}_i$ via a staging transformation with unit Jacobian \cite{Whitfield2007}. The
coefficients of the transformation are given in \ref{app:coeff}, leading to the following partition
function:
\begin{equation}
\begin{split}
  Z(\beta)&=\prod_{i=1}^{P} \int  \mathbf{d}\vec{x}_i \, \rho_0(\vec{x}_i, \vec{x}_{i+1}; \tau)
           \times \exp\left( -\tau \sum_{i=1}^P V(\vec{x}_i)\right) \\
          &= \prod_{i=1}^{P} \int  \mathbf{d}\vec{u}_i \, 
           \underbracket{
            \left(\frac{1}{2 \pi \sigma_i^2}\right)^{3N/2} 
             \exp\left( -\frac{\vec{u}_i^2}{2 \sigma_i^2}\right)
             }_{\text{staging coordinates}}
              \times \underbracket{\exp\left( -\tau \sum_{i=1}^P V(\vec{x}_i(\vec{u}_i))\right)}_{
              \text{external potential}
              }\,.
\end{split}
\end{equation}

One step remains in order to make the partition function isomorphic to the classical limit: 
the addition of conjugate momenta
$\vec{p}_i$ with corresponding faux masses $m_i$.
This transforms it to
\begin{equation}
\begin{split}
  Z(\beta) =& \prod_{i=1}^{P} \int  \mathbf{d}\vec{u}_i \, \left(\frac{1}{2 \pi \sigma_i^2}\right)^{3N/2} 
             \exp\left( -\frac{\vec{u}_i^2}{2 \sigma_i^2}\right)
              \times \exp\left( -\tau \sum_{i=1}^P V(\vec{x}_i(\vec{u}_i))\right) \\
              &\times C \int \mathbf{d} \vec{p}_i \, \exp\left( -\tau \frac{\vec{p}_i^2}{2 m_i}\right).
\end{split}
\end{equation}
This final transformation leads to a  partition function with 
the  effective classical Hamiltonian $ H^{\mathrm{(faux)}}$, which can be sampled via existing classical means 
\cite{Tuckerman1993, Whitfield2006, Whitfield2007}
\begin{equation}
  H^{\mathrm{(faux)}} = \sum_{i=1}^P \frac{\vec{p}_i^2}{2 m_i} + \frac{\vec{u}_i^2}{2 \sigma_i^2 \tau}
                        + \frac{V(\vec{x}_i(\vec{u}_i))}{P}.
\end{equation}

When adding nuclei, the adiabatic principle needs to be invoked and the temperature of
the electronic degrees of freedom increased by a factor of $\gamma$ relative 
to that of the nuclear degrees of freedom to reduce the number of beads required for 
convergence. \cite{Tuckerman1993, Whitfield2006, Whitfield2007, Jones2013b}

\section{QDO\_MD: a molecular dynamics software implementing Quantum Drude Oscillators} 

The QDO methodology has been implemented in QDO\_MD, a new software based on the
work of the Tuckerman and Martyna groups that includes a classical molecular 
dynamics engine, Car--Parrinello ab initio molecular dynamics (CPAPIMD) \cite{Kale2005}
using plane wave basis set, as well as a QM/MM capability with 
$N\log(N)$ scaling. QDO\_MD encompasses a number of advanced algorithms including 
parallel tempering \cite{Swendsen1986} and centroid path integral propagators for nuclear degrees 
of freedom \cite{Tuckerman2000}. The force field based part of the software has the standard 
force fields built in, targeting complex chemical and biological applications.
QDO\_MD is based on a number of algorithms developed by its principal authors, which are 
now widely accepted, including reversible reference system propagator (RESPA) \cite{Tuckerman1992}, 
Nos{\'e}--Hoover chains \cite{Martyna1992} and isothermal--isobaric ensemble \cite{Martyna1994} 
and allows simulation in various equilibrium statistical ensembles, perform energy and structure 
minimization using several techniques including steepest descent, conjugate gradient or direct 
inversion in the iterative subspace.
It is parallelised \cite{Collis2009} at several levels including classical force 
parallelisation, parallelisation over beads in path integral simulation, state and hybrid 
state/reciprocal space parallelisation of the electrostatic interactions and is aimed at 
distributed memory architectures using the Message Passing Interface (MPI) protocol \cite{MPI2015}.
QDO\_MD is undergoing an extensive refactoring and is currently available by contacting the authors. 
In the
near future, it will be placed on the National Physical Laboratory website and open
sourced on GitHub.

QDO\_MD has modular structure and easily allows extensions conforming to the accepted 
structure. In order to perform adiabatic path integral molecular dynamics with QDO 
(APIMD-QDO), a new simulation type keyword, \texttt{qdo\_pimd}, has been added. Two new atom types, 
\texttt{drude\_bead} and \texttt{drude\_center} specify the drudon. For generality, the Drude 
oscillator centres are treated as ghost atoms. 
Electrostatic interactions in QDO\_MD are calculated using smooth particle mesh Ewald (PME) 
method \cite{Essmann1995}, and in addition to standard 3D-periodic systems the code can 
handle the systems of reduced dimensionality using modified reciprocal space 
summation \cite{Martyna1999,Minary2002}. Although drudons do not introduce 
new types of interactions, special terms have been added between the bead and the centre 
in drudon to correct for terms necessary included in the reciprocal space part of the 
Ewald scheme. In addition to that, Coulomb regularisation has been added at short range 
as well as triexponential repulsion terms to correct for the missing exchange repulsion.
The pressure tensor, which is calculated in QDO\_MD from the atomic virial expression 
is currently being converted to a virial form based on heavy atom centres. 
Following the code convention, all interactions are splined. 

APIMD-QDO integration is done in the canonical ensemble using a velocity Verlet propagator 
with RESPA and due to separation of the nuclear and QDO degrees of freedom this implementation 
requires separate thermostats for the Drude and atomic degrees of freedom. The separation is controlled by the
adiabaticity parameter $\gamma$, with QDO degrees of freedom kept at elevated temperature. 
In order to remove large vacuum energy of the Drude subsystem, low variance harmonic 
staging virial estimator \cite{Whitfield2007} acting on QDO sites only has been implemented
to sample the instantaneous energy.
Since APIMD-QDO is incompatible with certain molecular systems, e.g., in the 
current implementation the QDO molecules cannot simultaneously have the holonomic constraints,
the program performs a set of checks before starting the simulation and logs the error if the
conditions are violated.

APIMD-QDO can handle mixed systems, where only a subsystem contains QDO with the rest treated 
classically.
Since in the APIMD-QDO algorithm all ingredients for the classical Drude oscillator are included,
QDO\_MD could also simulate the classical polarizable molecules. These, however, cannot be 
mixed with QDOs.

\section{QDO-water: an electronically-coarse grained model of water}
The first application of QDOs to complex systems was water \cite{Jones2013a, Sokhan2015a}. 
Water was chosen due to
its fundamental importance for life. Its properties have been studied for decades,
yet the essential physics underlying  many of its live-giving anomalies remains unresolved \cite{Angell2014}.

\begin{figure}[htp]
\centering
  \includegraphics[width=0.6 \textwidth]{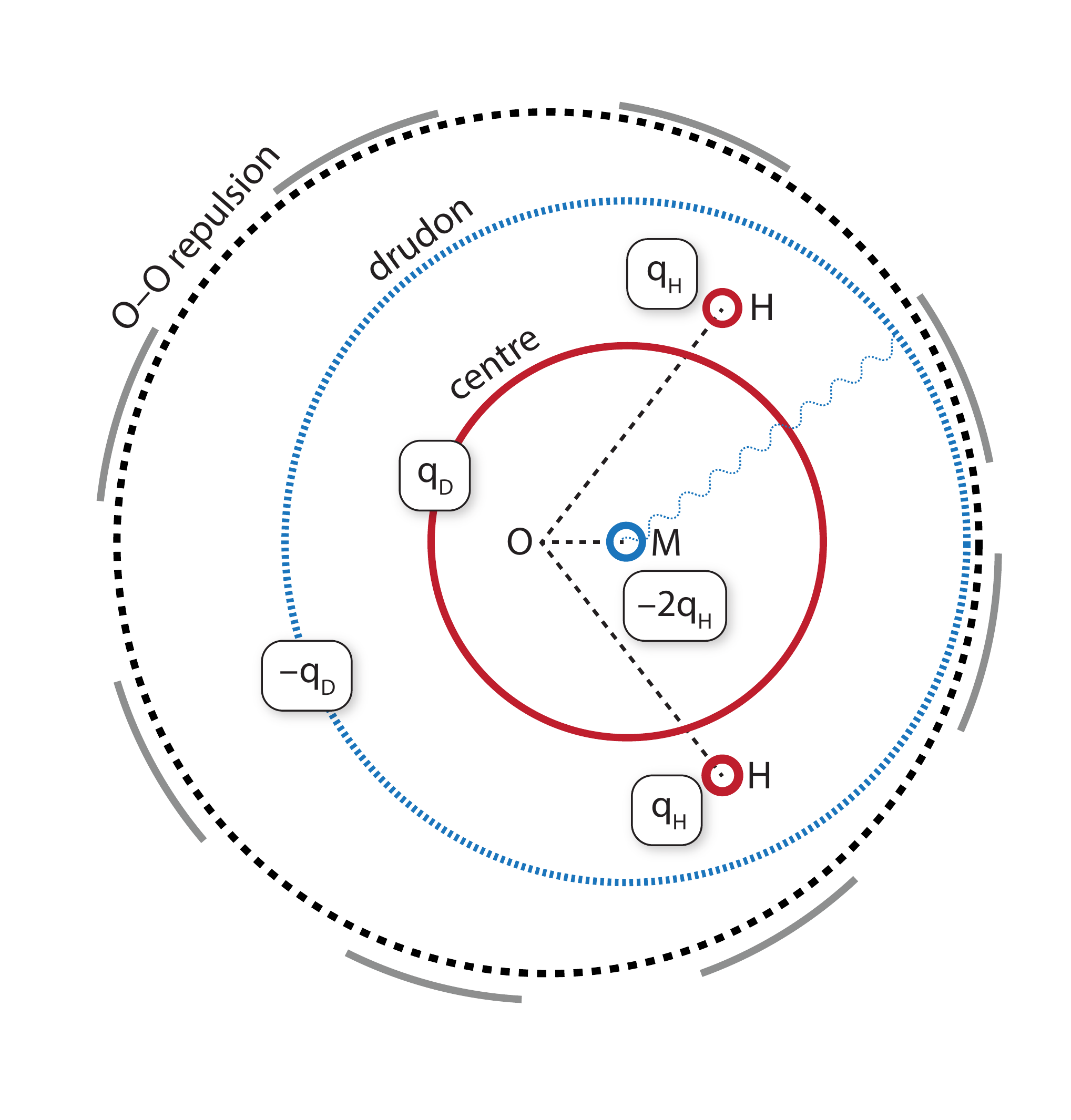}
\caption{Schematic of QOD-water.}
\label{fig:qdo-water}
\end{figure}

To construct a non-dissociative water model out of QDOs, one needs three elements: 
a rigid molecular frame with embedded point charges to replicate the lowest order
electrostatic moments, a QDO to replicate electronic responses and a short range, pairwise 
repulsion potential. These elements
are shown in Fig.~\ref{fig:qdo-water} with their respective
 parameters given in Table~\ref{table:qdo-water}.

\begin{table}[htp]
\centering
\begin{tabular}{cccc}
  \toprule
  Parameter & Value & Parameter & Value \\ 
     \cmidrule(l){1-2} \cmidrule(l){3-4}
  %%%%%%%%%%%%%%%%%%
  \multicolumn{2}{c}{\textbf{Molecular geometry}} & \multicolumn{2}{c}{\textbf{Coulomb damping}} \\
   $R_{\mathrm{OH}}$      &  $0.9572\,$\AA  & $\sigma_D = \sigma_H = \sigma_M$ & $0.1 \, a_0$ \\
   $\angle \mathrm{HOH}$  &  $104.52^\circ$ & $\sigma_C$                       & $1.2 \, a_0$ \\ 
  %%%%%%%%%%%%%%%%%%
  \multicolumn{2}{c}{\textbf{Ground state electrostatics}} & \multicolumn{2}{c}{\textbf{Short range repulsion}} \\
   $q_\mathrm{H}$    & $0.605 |e|$  & $\kappa_1$ & $613.3 \, E_h$  \\
   $R_{\mathrm{OM}}$& $0.2667\,$\AA & $\lambda_1$ & $2.3244 \, a_0^{-1}$  \\
   %%%%%%%%%%%%%%%%%%
   \multicolumn{2}{c}{\textbf{Quantum Drude Oscillator}} & $\kappa_2$ & $10.5693 \, E_h$  \\
   $m_{\mathrm{D}}$ & $0.3656 \, m_e$ & $\lambda_2$ & $1.5145 \, a_0^{-1}$ \\
   $\omega_{D}$ & $0.6287 \, E_h / \hbar$ & &  \\
   $q_\mathrm{D}$ & $1.1973 |e|$ & & \\
  \bottomrule
  \end{tabular}
\caption{The free parameters of QDO-water. $E_h \approx 27.211 \textrm{eV}$ is the Hartree energy, $a_0 \approx 0.5292 \textrm{\AA}$ is the 
        Bohr radius and $e \approx 1.60 \times 10^{−19}$ C is the electron charge}
\label{table:qdo-water}
\end{table}

In more detail, the frame is fixed in the
experimental molecular geometry, with charges and distances chosen to give an 
exact dipole moment (1.85 D) and a best fit to the quadrupole moment. The parameters of the QDO 
are fit using the relations in equation \eqref{eq:fitting} to the dipole and quadrupole polarisabilities
and the $C_6$ induced-dipole--induced-dipole dispersion coefficient. 
The repulsive potential is fit by calculating the 
interaction energy between two molecules using ab initio and the repulsion-free model and
fitting the difference to a double exponential $\kappa_1 e^{-\lambda_1 r} + \kappa_2 e^{-\lambda_2 r}$
\cite{Jones2013a, Sokhan2015a}. The Coulomb interaction between charges is regularised by
replacing each point charge $X$ by a Gaussian distribution of width $\sigma_{X}$.

Since QDO-water is parameterised from the properties of a single molecule and the dimer, 
condensed
phase properties are a prediction rather than a fitting target. The model was 
studied in the following environments: high pressure ice (ice II) \cite{Sokhan2015a},
liquid water \cite{Sokhan2015a}, steam \cite{Sokhan2015a}, the liquid-vapour interface \cite{Cipcigan2015} 
and supercritical water \cite{Sokhan2015b}. The agreement with experiment was good,
with densities within few percent of experiment. 
The model predicts the temperature of maximum
density of $5.5(2)^\circ\textrm{C}$ (compared to the experimental value of $3.98^\circ\textrm{C}$ \cite{Wagner1999})
and the critical point of $\{649(2) \textrm{K}, 0.317(5) \textrm{g/cm}^3\}$
(compared to the experimental value of $\{647.096 \textrm{K}, 0.322 \textrm{g/cm}^3\}$ \cite{Wagner1999}).
 It also accurately tracks
the dielectric constant, enthalpy of vaporisation and surface tension between 300 K and
600 K \cite{Sokhan2015a}.

QDO water also provided insights into the local structure at the liquid-vapour interface
\cite{Cipcigan2015} and
in supercritical water \cite{Sokhan2015b}. At the liquid-vapour interface the orientation of the water molecules
is influenced by an intrinsic asymmetry in the way water molecules hydrogen bond. When
losing a hydrogen bond, molecules prefer to lose it on the side of the oxygen (an acceptor bond)
rather than on the side of the hydrogen (a donor bond) \cite{Cipcigan2015}. This leads to a surface with
more oxygen than hydrogen atoms, which is negatively charged.

In supercritical water \cite{Sokhan2015b}, the dipole moment was discovered to be a reliable fingerprint of 
the liquid-gas transition \cite{Sokhan2015b}. The phase diagram locus where the heat capacity is maximum
coincides with a cusp in the dipole moment as a function of density. This cusp separates
liquid-like from gas-like scaling, tracking the Widom line. 
This conclusion is general to any polar fluid, showing
the importance of polarisation in characterising highly heterogeneous materials. 

\section{Conclusion}
This paper summarises our recent progress in significantly enhancing the transferability of 
force fields describing complex systems of insulators
 by coarse graining the electrons of a molecular moiety to
a single quantum harmonic oscillator, known as a Quantum Drude Oscillator. This
technique generates, naturally and to all orders, many-body forces such as dispersion, 
polarisation and 
mixed interactions beyond the dipole limit. These interactions are important
in modelling heterogeneous environments such as those occurring in biological systems,
where the symmetry of interactions depends strongly on environment.
To demonstrate the application of the model, we highlight the successful construction
of a transferable model of a water molecule. The model is parametrised using only
the properties of a single molecule and those of a dimer yet has excellent predictive 
power across the phase diagram, from high pressure ice to supercritical water. 
To encourage the adoption of the method, we present its
first implementation in the general purpose molecular dynamics software QDO\_MD.

\section{Acknowledgements}
This work was supported by the NPL Strategic Research programme. 
FSC acknowledges the Scottish Doctoral Training Centre in Condensed Matter Physics, 
NPL and EPSRC for funding under an Industrial CASE studentship. 
We acknowledge use of Hartree Centre resources in this work, 
including those from the Intel\textsuperscript{\textregistered} Parallel Computing Centre.

\appendix

\section{Coefficients}
\label{app:coeff}
For a quantum Drude oscillator of mass $m$, charge $q$ and frequency $\omega$, the
coefficients defined in the text have the following values,
\begin{equation}
\begin{split}
\alpha_P(\beta) &= \frac{m \omega}{2 \hbar \sinh(f)},\\
\lambda_P(\beta) &= \frac{2 m \omega \tanh(f/2)}{\hbar}, \\
f &= \frac{\beta \hbar \omega}{P}
\end{split}
\end{equation}

The staging transformation uses the following coefficients:
\begin{equation}
\begin{split}
  \vec{u}_1 &= \vec{x}_1, \\
  \vec{u}_i &= \vec{x}_i - \vec{x}_i^*, \\
  \vec{x}_i^* &= \frac{\sinh(\tau \hbar \omega)}{\sinh[i \tau \hbar \omega]} \vec{x}_1
                 + \frac{\sinh((i-1)\tau \hbar \omega)}{\sinh(i \tau \hbar \omega)} \vec{x}_{i+1}\,, \\ 
  \sigma_1^2 &= \frac{\hbar}{2 m \omega \tanh(\beta \hbar \omega / 2)}\,, \\
  \sigma_i^2 &= \frac{\hbar \sinh[(i-1)\tau \hbar \omega] \sinh(\tau \hbar \omega)}{m \omega \sinh[i \tau \hbar \omega]}\,.
\end{split}
\end{equation}

\bibliographystyle{elsarticle-num} 
\bibliography{jcp-cambridge}

\end{document}